\documentclass[11pt]{article}
\usepackage{theorem}
\usepackage{amsmath}
\usepackage{amsfonts}
\usepackage{amssymb}
\usepackage{euscript}
\usepackage{hyperref}
\hypersetup{colorlinks=true}
\newcommand{\R}{\mathbb{R}}

\newcommand{\fraX}{\mathfrak{X}}

\newcommand{\p}{\partial}
\newcommand{\del}{\delta}
\newcommand{\leven}{\Lambda^{\kern -11pt 0}}
\newcommand{\lodd}{\Lambda^{\kern -11pt 1}}

{\theoremstyle{break} \theorembodyfont{\slshape} }

{\theoremstyle{break} \theorembodyfont{\rmfamily} \newtheorem{example}{Example}[section]}
\begin{document}
\begin{titlepage}
\begin{center}
{\large \bf On a set of Grassmann-valued extensions\\of systems of ordinary differential equations}
\vskip 4em
{\large M. L\'egar\'e\;$^\ast$}
\\
\vskip 1em
Edmonton, Alberta, Canada\\
\end{center}
\vskip 2em

\begin{abstract}
 
Formulations of some Grassmann-valued systems of ordinary differential equations invariant under (infinitesimal) supersymmetry transformations, including $N$-superspace extended types, are reviewed and discussed, with use of superfields. Different new examples are shown, and some aspects on methods for obtaining solutions are offered. Notions inspired from Darboux theory are considered for some general polynomial systems, involving Grassmann variables. 

\medskip
\noindent\sloppy{{\bf MSC(2010)} : 81T60,17B81}

\noindent\sloppy{{\bf PASC(2010)} : 02.30.Hq,11.30.Pb}

\medskip
\noindent\sloppy{{\bf Keywords} :  system of ordinary differential equations, supersymmetric systems}

\end{abstract}
\vfill
$^\ast$ \small{For correspondence with the author, please use : mlegare@ualberta.ca  \newline or mrtinlegare@gmail.com}

\end{titlepage}

\pagebreak

\tableofcontents
\noindent
\section{Introduction}

Grassmann-valued extensions of different systems and models in various dimensions have been largely exhibited and studied. Among many articles, only a few on this subject are mentioned in this paper, see for instance references \cite{BGHW,FKT,AH,LY,GH}. It is noted that invariance under supersymmetry transformations is often obeyed by many extended systems. The original system would be retrieved from an extension or supersymmetric version by setting the supplementary or additional variables to zero.

In this article, a focus will be put on generating Grassmann-valued extensions of systems of o.d.e.s. The approach will be mainly based on the introduction of superfields, either even or odd. Invariance with respect to (infinitesimal) supersymmetry transformations (associating anti-commuting behaviour with fermionic aspects) will be sought. Certain aspects will be reviewed and discussed. To our knowledge, new supersymmetric extensions are exhibited. Also, a set of notions inspired from concepts of Darboux theory are developed and explored within the Grassmann variables setting.

Many evolution equations in (1+1) dimensions have been extended to supersymmetric versions involving solely supersymmetric transformations associated with one (spatial) bosonic dimension, whether it is for example the supersymmetric Korteweg - de Vries (KdV) equation \cite{Ma} (as well as $N = 2, 4$ extended forms), the supersymmetric Nonlinear Schr\H{o}dinger (NLS) equation \cite{RK}, or the supersymmetric Harry Dym equation \cite{BDP}. Supersymmetry representations in different contexts have been shown, such as in supersymmetric mechanics \cite{BK,KKLNS}.

A complete set of scalar $\lambda$ - homogeneous (where $\lambda > 0$) $N = 1$ supersymmetric evolutionary integrable equations, which are labelled ``nontrivial extensions'', has been provided in \cite{TW}. The term ``nontrivial'' refers to a (truly) coupled system expressed in terms of components of the superfield, see \cite{TW} for details. In the following,  a trivial extension of a system would relate to an extended system where the original equation(s), then written in terms of Grassmann-valued (even or odd) components (of superfield(s)), are unaffected by the other (even or odd) fields additions. In the trivial case, the original equations written in terms of Grassmann-valued versions of the original variables would be uncoupled from the other even or odd variables.

It is mentioned that reductions by symmetry of certain higher dimensional supersymmetric systems, for example using invariance under translations with respect to certain variables, could lead to supersymmetric extensions of ordinary differential equations (abbreviated o.d.e.s) or systems of o.d.e.s.
 
 The plan of this paper includes a second section, where an extension of certain Darboux theory notions \cite{Ll,LVZ,Zh} is suggested for planar systems in superspace. Odd or fermionic superfield extensions are also discussed and $N=1$ superspace versions proposed for certain systems of o.d.e.s. Section 3 explores the case of $N$ - superspace extensions of different systems, with some examples such as a simple polynomial equation, a H\'enon - Heiles system for $N = 2$, and a $N=3$ extension of the Euler - Arnold equations. In section 4, discussions are offered on some methods of solution. A method is based on a Grassmann algebra basis together with the help of series solutions of systems of o.d.e.s. Another uses supersymmetry transformations on known seed solutions. A third approach involves polynomial systems with modified products called non - associative (more precisely not - associative) products \cite{Sch}, for which a series solution approach is described. A conclusion follows in section 5.

\section{N=1 extensions and some Darboux theory aspects on Grass\-mann-valued systems of o.d.e.s}

In this section, some systems of o.d.e.s possessing supersymmetry invariance and simple generalizations of a planar dynamical system with Grassmann variables are considered.

\subsection{Supersymmetric systems and planar superspace}

Let us consider on superspace \cite{Fre,Rog} $\R_{S[L 0]} \times \R_{S[L 1]}$ (see \cite{Rog} for notation, but in this article $\R_{S[L 0]}$ is also denoted by $^0\Lambda_L$ and $\R_{S[L 1]}$ by $^1\Lambda_L$), coordinates $t \in \R_{S[L 0]}$ and $\theta \in \R_{S[L 1]}$. The label $L$ could be finite or infinite ($L \rightarrow \infty$) corresponding to a finite or infinite  dimensional  Grassmann algebra. But in the following, unless otherwise mentioned, $L$ will be taken as infinite. However, for brevity, the short notation $^0\Lambda$ and $^1\Lambda$ will be used to indicate respectively even and odd spaces of the finite or infinite Grassmann algebra. An (even) scalar superfield is denoted $X$ :  $X(t,\theta) = x(t) + \theta \xi(t)$, where $x \in$$^0\Lambda$ and $\xi \in$$^1\Lambda$ form together what will be called a planar superspace. A substitution of $x$ and $\xi$ in the above definition of $X$, that is : $Y(t,\theta) = \xi(t) + \theta x(t)$, would lead to an odd scalar superfield, which will also be used in what follows.

Use of scalar superfields will be made, these descriptions could lead to what are sometimes called B-supersymmetrization (see for example \cite{CTG}), which have had some interests (for instance \cite{BB,FS}). However, it seems for many cases given as Grassmann-valued systems, that they provide, when expanded in terms of a basis of a suitable Grassmann algebra, the same original equations at the zero (of a nilpotent variable ``Taylor'' expansion) level or body level (\cite{Rog}).

In the article of Heumann and Manton \cite{HM}, the variable $t$ could belong to the body of $^0\Lambda$, a setting that will be adopted below in a set of situations (for aspects on Grassmann analytic continuations, see for example \cite{Rog}, sections 4.2 and 4.3). Inverses are supposed to exist when needed (for even variables).

With $F(X)$ being a polynomial in the superfield $X(t,\theta)$, one can introduce a ``planar'' differential equation for $x$ and $\xi$ (see for instance \cite{Fre}) :
\begin{equation}
\dot X = F(X)
\end{equation}
where ``$\;\dot{}\;$'' indicates a derivative with respect to the $t$ variable. This system can be seen as a generalization to Grassmann variables of planar dynamical systems such as those studied in \cite{Ll,LVZ,Zh}, since :
\begin{equation}\label{planar-super}
\dot x = F(x), \quad \dot \xi = F'(x) \xi
\end{equation}
where $F'(x)$ denotes the derivative with respect to $x$ of $F(x)$.

Introducing some elements inspired from Darboux theory (see for example \cite{Zh}), one defines a (polynomial) vector field ($\mathcal{X}$) :

\begin{equation}
\mathcal{X} = F(x) \frac{\p}{\p x} + (F'(x) \xi) \frac{\p}{\p \xi}
\end{equation}
If the polynomial $f(x,\xi)$ obeys to :
\begin{equation}
\mathcal{X} (f) = \kappa f
\end{equation}
where $\kappa(x,\xi)$ is also a polynomial in $x$ and $\xi$, called here Grassmann cofactor, then $f(x,\xi)$ corresponds to what could be seen as a Grassmann Darboux polynomial. Let us note that $\xi$ and $F(x)$ are themselves Grassmann Darboux polynomials with each having $\kappa(x,\xi) = F'(x)$ as cofactor. Thus $I(x,\xi) = \frac{\xi}{F(x)}$, will be called a  Grassmann rational first integral of this system (see for instance Zhang (35) for non-extended aspects), that is $\dot I = 0$, or $\mathcal{X}(I) = 0$. A solution to the system (\ref{planar-super}) can be found with :
\begin{equation}
\int \dfrac{dx}{F(x)} = t + C_1,\; \text{and}\; \xi(t) =  C_2 \exp\left ({\int F'(x)\; dt}\right )
\end{equation}
where $C_1 \in$ $^0\Lambda$ and $C_2 \in$ $^1\Lambda$ are constants of integration.

The system (\ref{planar-super}) is invariant under (supersymmetry) transformations (see for example \cite{Fre,HM,AGMT,RCG,CRG,Man}). Infinitesimally, one has :
\begin{equation}
\del X  = \epsilon Q X
\end{equation}
or component-wise :
\begin{equation}
\del x = \epsilon \xi, \quad \del\xi = -\epsilon \dot x 
\end{equation}
where $\epsilon$ is a $^1\Lambda$ constant parameter, $X = x(t) + \theta \xi(t)$, and $Q$ is the operator given by $Q = \left ( \frac{\p}{\p \theta} + \theta \frac{\p}{\p t} \right )$. ($t$ transformed to $-t$ brings a definition of $Q$ used in other articles.)

\subsection{Planar Grassmann-valued systems}

Let us propose a system of two dependent variables $x(t)$ and $\xi(t)$,  which is not necessarily invariant under supersymmetry (or Grassmann-valued) transformations. The following polynomial planar Grassmann-valued system :
\begin{equation}\label{planar}
\dot x  = P(x,\xi), \quad \dot \xi = Q(x,\xi)
\end{equation}
is discussed below. As above, $x \in$ $^0\Lambda$ and $\xi \in$ $^1\Lambda$, also $P(x,\xi)$ and $Q(x,\xi)$ are both polynomials in $x$ and $\xi$. Because of the nilpotency of $\xi$, the system (\ref{planar}) reduces to :
\begin{equation}
\dot x = p(x) + q(x) \xi,\quad  \text{and} \quad \dot \xi = \hat p(x) + \hat q(x)  \xi 
\end{equation}
where $p(x), \hat q(x) \in$$^0\Lambda$, and $q(x), \hat p(x) \in$$^1\Lambda$ are polynomials in $x$. The supersymmetric system (\ref{planar-super}) is a particular case with $p(x) = F(x), q(x)= 0, \hat p(x) = 0$, and $\hat q(x) = F'(x)$. An associated vector field $\mathcal{X}$ to equation (\ref{planar}) is written as :
\begin{equation}
\mathcal{X} = P(x,\xi) \frac{\p}{\p x} + Q(x,\xi) \frac{\p}{\p \xi} = (p(x) + q(x) \xi) \frac{\p}{\p x} + (\hat p(x) + \hat q(x) \xi) \frac{\p}{\p \xi}
\end{equation}

Analogously to the ordinary, or non-extended (or non-Grassmannian) case, Grassmann Darboux polynomials $f(x,\xi)$ would be defined, similarly to above, as those satisfying the condition : $\mathcal{X} f(x,\xi) = \kappa(x,\xi) f(x,\xi)$.

The qualifying term ``Grassmann'' will not be necessarily added in the following when the context suggests it.

\begin{example}[Darboux polynomials]
Let us pick $f(x,\xi) = \xi$. $f$ will be called a Grassmann Darboux polynomial, if $\mathcal{X}(\xi) = \hat p(x) + \hat q(x) \xi = \kappa(x,\xi) \xi$, where $\kappa(x,\xi) = \kappa_1(x) + \kappa_2(x) \xi$, since again $\xi$ is a nilpotent variable. Thus $\kappa_1(x) \xi = \hat p(x) + \hat q(x) \xi$, which is obeyed if : $\kappa_1(x) = \hat q(x)$ and $\hat p(x) = 0$.
\hfill $\ddagger$\end{example}

\begin{example}[First integral]
Let us explore the expression $I(x,\xi) = \xi h(x)$, where $h(x) \in$ $^0\Lambda$ is at first an arbitrary function of $x$. The quantity $I$ would be defined as a Grassmann ``first integral'' if $\mathcal{X} f(x,\xi) = 0$ is satisfied. This latter condition requires the following :
\begin{equation}
\hat p(x) h(x) + (p(x) h'(x) + \hat q(x) h(x)) \xi = 0
\end{equation}

This relation is respected if $\hat p(x) h(x) = 0$, and $p(x) h'(x) + \hat q(x) h(x) = 0$.
Setting $\hat p(x) = 0$ and $\hat q(x) = - p(x) \frac{h'(x)}{h(x)}$, allows to obtain a ``planar'' system with such first integral $I$, which has the form :
\begin{equation}
\dot x = p(x) + q(x) \xi, \quad \dot \xi = - p(x) \frac{h'(x)}{h(x)} \xi,
\end{equation}
where the term $-p(x) \frac{h'(x)}{h(x)}$ is asked to be a polynomial, denoted $r(x)$, while $p(x)$ is also set to be a polynomial.

An arbitrariness in the function $h(x)$ permits to have : $\frac{h'(x)}{h(x)} = - \frac{r(x)}{p(x)}$, and therefore, $h(x)$ could be an expression such as : 
\begin{equation}
h(x) = A \exp\left (- \int \frac{r(x)}{p(x)}\;dx \right ),
\end{equation}
where $A \in$$^0\Lambda$, is a constant of integration. 

A simple calculation for the system : $\dot x = x^2 + \alpha \xi, \dot \xi = -x^3 \xi$, where $\alpha \in$$^1\Lambda$, is a constant, with $p(x) = x^2$, leads to $h(x) = A \exp \left ( x^2/2 \right )$, and a first integral $I(x,\xi) = A \xi \exp \left ( x^2/2 \right )$, for the sought function $r(x) = -x^3$. There, the function $\exp \left ( x^2/2 \right )$ can be labelled as a (Grassmann) Darboux function.
\hfill $\ddagger$\end{example}

In the above, systems (on $\R^{1,1}$) where $p(x) + q(x) \xi \neq 0$, and  / or $\hat p(x) + \hat q(x) \xi \neq 0$ are considered.

It is noted that a component form of the superfield equations (also called ``bosonization'' in certain articles (see for example \cite{GL,R}) will provide systems of (first-order) o.d.e.s which can then be studied, possibly with Darboux theory. For instance, the system $\dot X = X^2$, with $X = x + \theta \xi$, can be written as : $\dot x = x^2, \dot \xi = 2 x \xi$, where $x \in{}^0\Lambda$ and $\xi \in{}^1\Lambda$. If $L=2$ with basis of generators $e^1,e^2$, then one can decompose the variables as follows: $x = x_0 + x_{12} e^1e^2, \xi = \xi_1 e^1 + \xi_2 e^2$, where $x_0, x_{12}, \xi_1, \xi_2$ are real - valued functions of the independent variable $t$. A system of o.d.e.s for these real dependent variables would be derived. A set of solutions of the (above) Grassmann-valued system can be obtained.

\subsection{Odd or fermionic superfield extentions}

Most used in building supersymmetric extensions of systems, odd (or fermio\-nic) superfields are shown in this section with a few examples of interest. Only $N=1$ superspaces are discussed in this section.

\begin{example}[reduced KdV equation]
A well known ($N=1$) supersymmetric extension of the KdV equation is reviewed. This is an integrable case (see \cite{Ma},  a $x \rightarrow -x$ transformation to accommodate the below operator $Q$ has been used) :
\begin{equation}
u_t = u_{xxx} - 6 u u_x - 3 \xi \xi_{xx}, \quad \xi_t = \xi_{xxx} - 3(u \xi)_x,
\end{equation}
where $u = u(x,t) \in$ $^0\Lambda$ and $\xi = \xi(x,t) \in$$^1\Lambda$. An invariance with respect to translations along $t$ leaves $u = u(x)$, and $\xi = \xi(x)$, and gives rise to the following system of o.d.e.s :
\begin{equation}
\p_x \left ( u_{xx} - 3u^2 - 3 \xi \xi_x \right )  = 0, \quad \p_x \left (\xi_{xx} - 3 u \xi \right ) = 0.
\end{equation}
This reduced system is left unchanged by the supersymmetry transformations : $\del u = - \epsilon \xi_x, \quad \del\xi = \epsilon u$, also written as $\del Y  = \epsilon Q Y$, acting on $Y = \xi + \theta u$, where $Q = \left ( \frac{\p}{\p \theta} + \theta \frac{\p}{\p x} \right )$. With $D = \left ( \frac{\p}{\p \theta} - \theta \frac{\p}{\p x} \right )$, $D^2 =- \p_x$, and one can use $DY$ to retrieve the suitable $u^2$ term of the non-extended equation for the dependent variable $u$ at the $\theta$ level. One can write :
\begin{equation}
D^2(D^4 Y - 3Y DY) = 0,
\end{equation}
which has common traits with supersymmetric extensions of KdV equation.
\hfill $\ddagger$\end{example}

\begin{example}
This system has been given an extension in subsection 2.1 with even superfield.  A Grassmann-valued extension of the polynomial o.d.e. (with $x$ real valued in the non-extended version) :
\begin{equation}
\dot x =  F(x)
\end{equation}
can be considered with an odd homogeneous superfield : $Y =  \xi + \theta x$. It follows that : $DY = x - \theta \dot \xi$. If one sets for simplicity  $F(x) =  x^n$, then an extension can be formulated as :
\begin{equation}
\dot Y  = Y (DY)^{n-1},
\end{equation}
with  for $n \geq 2$, $(DY)^{n-1} = (x-\theta \dot \xi)^{n-1} = x^{n-1}  - \theta (n-1) x^{n-2} \dot \xi$.
For a finite degree polynomial $F(x) = \sum_{j=0}^n a_j x^j$, with coefficients $a_j$, one could have the extension :
\begin{equation}
\dot Y = Y (\sum_{j=1}^n a_j (DY)^{j-1}) + a_0 \theta
\end{equation}
which differs, once written in component form, from the system (\ref{planar-super}) .
\hfill $\ddagger$\end{example}

\subsection{Simple Grassmann-valued extensions of first-order differential systems}

In the following, the variables $x_1, x_2, ...$ are defined as real valued functions in non-Grassmannian versions, and belong to $^0\Lambda$ in Grassmannian extensions.

Given a (polynomial) differential system :
\begin{equation}
\dot{\vec x} = \vec P(\vec x),
\end{equation} 
where $\vec x = (x_1, x_2, ... , x_p)^T$, with $(...)^T$ as the transpose, and $x_i, i = 1, 2, ... ,p$, each being a function of ``$t$'', $x_i = x_i(t)$, and where $\vec P = (P_1, P_2, ... , P_p)^T$ with $P_i, i = 1, ,2 , ... , p$, being polynomials in $x_j, j = 1, 2, ..., p$.

One can offer a simple extension of such differential systems involving supersymmetry transformations. A substitution of $x_i$ by the even superfields $X_i = X_i(t,\theta) = x_i(t) + \theta \xi_i(t)$, where $x_i \in$$^0\Lambda$, and $\xi_i \in$$^1\Lambda$ for each $i= 1, ... , p$, leads to :
\begin{equation}
\dot{\vec X} = \vec P(\vec X),
\end{equation}
and in detail :
\begin{equation}
\dot x_i = P_i(x_j), \quad \dot \xi_i = \frac{\p P_i}{\p x_j} \xi_j 
\end{equation}
with $i,j = 1, ... ,p$ with sum over repeated index $j$. A corresponding vector field ($\mathcal{X}$) could be defined :
\begin{equation}
\mathcal{X} = P_i(x) \frac{\p}{\p x_i} + \left (\frac{\p P_i}{\p x_j} \xi_j \right ) \frac{\p}{\p \xi_i}
\end{equation}
with summation over repeated indices. Adaptations of definitions of Darboux polynomials, with cofactors, and first integrals can be brought in this setting.
Hence, $f(X_i) = f(x_i,\xi_i)$ is a Darboux polynomial if : $\mathcal{X} f(x_i,\xi_i) = \kappa(x_i,\xi_i) f(x_i,\xi_i)$, where $\kappa(x_i,\xi_i)$ would be called a cofactor, and the equation $\mathcal{X} I(x_i,\xi_i) = 0$, being satisfied would bring $I(x_i,\xi_i)$ as a first integral of the system. In a similar fashion to non-extended Darboux polynomials, the vanishing of polynomials $f(x_i,\xi_i)$ would lead to objects reminiscent of ``invariant surfaces'' in the $\{(x_i,\xi_i), i = 1,...,p\}$ - space. 

\begin{example}[Fermionic superfields]
An extension of the (simple) first-order system :
\begin{equation}
\dot x_1 = x_1 x_2, \quad \dot x_2 = x_2^2,
\end{equation}
is sought. A set of odd scalar superfields : $Y_1(t,\theta) = \xi_1(t) + \theta x_1(t)$ and $Y_2(t,\theta) = \xi_2(t) + \theta x_2(t)$ would provide a Grassmann-valued extension in the form :
\begin{equation}
\dot Y_1 = Y_1 (DY_2), \quad \dot Y_2 = Y_2 (DY_2),
\end{equation}
where $D = \left ( \frac{\p}{\p \theta} - \theta \frac{\p}{\p t} \right )$, which anti-commutes with the operator $Q$ previously defined ($\{Q,D\} = QD + DQ = 0$).
The extended system can be written in components as : 
\begin{equation}\label{super-odd-2}
\dot x_1 = x_1 x_2 + \xi_1 \dot \xi_2, \; \dot \xi_1 = \xi_1 x_2, \; \dot x_2 = x_2^2 + \xi_2 \dot \xi_2, \; \dot \xi_2 = \xi_2 x_2,
\end{equation}
which reduces to :
\begin{equation}\label{super-odd-2-red}
\dot x_1 = x_1 x_2 + \xi_1 \xi_2 x_2, \; \dot \xi_1 = \xi_1 x_2, \; \dot x_2 = x_2^2, \; \dot \xi_2 = \xi_2 x_2,
\end{equation}
inserting the equation for $\xi_2$ in the equation for $x_1$ and $x_2$ of the system (\ref{super-odd-2}), and using $\xi_2$ nilpotency.

It can be verified that $\frac{\xi_1}{x_2}$ and $\frac{\xi_2}{x_2}$ are first integrals (as defined previously) of the above system (\ref{super-odd-2-red}), belonging to the kernel of the corresponding vector field ($\mathcal{X}$). However, the expression $\frac{x_1}{x_2}$ is a first integral of the (non-extended) original system, but is not of the extension (\ref{super-odd-2-red}). Instead, the following expression forms another first integral of the extended system :
\begin{equation}
I = \left (\frac{x_1 - \xi_1\xi_2}{x_2} \right )
\end{equation}
\hfill $\ddagger$\end{example}

\begin{example}[First-order system of reduced KdV]\label{first-order-kdv}

It is pointed out that the time translation invariant KdV equation : $u_{xxx} = 6u u_{x}$, could be rewritten as a first-order system of o.d.e.s, for instance :
\begin{equation}
{x_1}_x = x_2, \quad {x_2}_x = x_3, \quad {x_3}_x = 6 x_1 x_2,
\end{equation}
using a correspondence : $x_1 = u$. One can define $ \vec x = [x_1,x_2,x_3]^T$ to express the system simply as :
\begin{equation}\dot {\vec x} = A \vec x + \vec \Pi(\vec x), \quad A = \bmatrix 0 & 1 & 0 \\ 0 & 0 & 1 \\ 0 & 0 & 0 \endbmatrix
\end{equation}
with the homogeneous quadratic product $\vec \Pi(\vec x) = [0,0,6x_1x_2]^T$. A discussion of a non-associative description of this system could be proposed (see example \ref{kdv-na}). Nevertheless, supersymmetric extensions could be devised using either even superfields : $X_i = x_i + \theta \xi_i, i = 1,2,3$, or odd superfields : $Y_i = \xi_i + \theta x_i, i = 1,2,3$, along with insertion for the odd case of the operator $D$ in order to recover the non-extended system of equations (when $\xi_i = 0, i = 1,2,3$) with the independent variable $x$ playing the role of $t$. An explicit extended form is given in example \ref{kdv-na} with even scalar superfields.
\hfill $\ddagger$\end{example}

\begin{example}[System with 2 even superfields]

With $X_i = x_i + \theta \xi_i, i = 1,2$, where $x_i \in$$^0\Lambda$ and $\xi_i \in$$^1\Lambda$ for $i = 1,2$, one can propose the following ``nontrivial'' extension of a simple linear system ($\dot x_1= \alpha x_1, \dot x_2 = x_2$) :

\begin{equation}
\dot X_1 = \alpha X_1 + (DX_1)(DX_2), \quad \dot X_2 = X_2,
\end{equation}
where $\alpha \in \R$ is a constant. In components, one finds :
\begin{equation}\label{2-super-nontrivial}
\dot x_1 = \alpha x_1 + \xi_1\xi_2, \quad \dot \xi_1 = \alpha \xi_1 - \alpha \xi_2 x_1 + \xi_1 x_2, \quad \dot x_2 =  x_2. \quad \dot \xi_2 =  \xi_2  
\end{equation}

For any $\alpha \in \R$, a first integral can be given as : $\frac{\xi_2}{x_2}$. For $\alpha = 0$, one can verify that : $\frac{\xi_1}{e^{(x_2)}}$, is also a first integral for the above  system (\ref{2-super-nontrivial}). 
\hfill $\ddagger$\end{example}

\begin{example}[Three-wave resonant interactions extension]\label{twri}

Using real - valued functions $x_a, a = 1,...,6$, and the complex - valued functions $u_1 = x_1 + i x_2, u_2 = x_5 + i x_6$, and $u_3 =  x_3 + i x_4$, one can express the three-wave resonant interactions : $\dot u_1 = - u_2 u_3, \dot u_2 = u_1 u^\ast_3, \dot u_3 = u_1 u^\ast_2$ (see for instance \cite{ZM,GM,HBC}) as :
\begin{alignat}{2}\label{real-equations}
\dot x_1 &= -(x_3x_5 -x_4x_6)  &\quad \dot x_2 &= - (x_4x_5 + x_3x_6) \nonumber\\
\dot x_3 &= (x_1 x_5 + x_2 x_6) & \dot x_4 &= (x_2 x_5 - x_1 x_6) \\
 \dot x_5 &= (x_1x_3 + x_2x_4) & \quad \dot x_6 &= (x_2 x_3 -x_1 x_4) \nonumber
\end{alignat}

A supersymmetric extension on $N=1$ superspace with even scalar superfields : $X_a , a = 1,...,6$, substituting respectively the variables $x_a, a = 1,...,6$, allows to have invariants (first integrals) such as :
\begin{equation}
X_1^2 + X_2^2 + X_5^2 + X_6^2, \quad \text{and} \quad X_3^2 + X_4^2 - X_5^2 -  X_6^2
\end{equation}
respectively related to the non-extended invariants : $|u_1|^2 + |u_2|^2$ and $|u_3|^2 - |u_2|^2$. Let us note that the addition to the extended equations from (\ref{real-equations}) of terms such as : $(DX_a)(DX_b)$ could bring ``non-trivial'' extensions of the three-wave resonant interactions.

However, use of $N=1$ superspace odd scalar superfields : $Y_a = \xi_a + \theta x_a, a= 1,...,6$ can also be involved, with substitution of the variable $x_a$ by the superfield $Y_a$ for all $a = 1,...6$, and suitable insertion of terms $Y_a DY_b$ replacing the quadratic expressions $x_ax_b$, $a,b = 1,...,6$. First integrals or invariants of the non-extended system are not necessarily preserved by these changes.

As example of an extension for the equation of $x_1$ of (\ref{real-equations}), one writes :
\begin{equation}
\dot Y_1 = - \frac{1}{2}(Y_3DY_5 - Y_4DY_6 + (DY_3)Y_5 - (DY_4)Y_6),
\end{equation}
which in components provides :
\begin{align}
\dot \xi_1 &= -\frac{1}{2} (\xi_3x_5 - \xi_4 x_6 + x_3 \xi_5 - x_4 \xi_6)\\ 
\dot x_1 & = -(x_3x_5 - x_4x_6) - \frac{1}{2}(\xi_3\dot \xi_5 - \xi_4 \dot \xi_6 + \xi_5 \dot \xi_3 - \xi_6 \dot \xi_4) \nonumber
\end{align}
\hfill $\ddagger$\end{example}

\begin{example}[Darboux - Halphen system extension]\label{d-h}

Another known system of first - order differential equations can be dressed with a supersymmetric extension. The Darboux - Halphen system (see \cite{CGR} and references therein) is given by : $\dot x_1 = x_1(x_2 + x_3) - x_2x_3$, $\dot x_2 = x_2 (x_3 + x_1) - x_3x_1$, $\dot x_3 = x_3 (x_1 + x_2) - x_1x_2$, where $x_i, i = 1,2,3$, are real valued functions (see for instance \cite{CGR}).

A supersymmetric version of this quadratic system can be obtained with odd superfields : $Y_i = \xi_i + \theta x_i$, $i = 1,2,3$, via the following equations succinctly written :
\begin{equation}\label{s-darboux-halphen}
\dot Y_i = Y_i (DY_j + DY_k) - Y_jDY_k
\end{equation}
with $i=1, j=2, k=3$, and cyclic permutations of the three indices.

An infinite set of Grassmann-valued (supersymmetric) extensions can be built, for instance with terms of the kind :
\begin{equation}
a Y_i(DY_j + DY_k) + (1-a) (DY_i)(Y_j + Y_k) - (bY_jDY_k + (1-b) (DY_j) Y_k)
\end{equation}
replacing the right hand side of equation (\ref{s-darboux-halphen}), where $a$ and $b$ are real parameters. More general supersymmetric versions can be constructed. 

Let us also mention that generalized Lotka - Volterra systems can be supersymmetrically extended in a similar fashion, as in previous examples \ref{twri} and \ref{d-h}.
\hfill $\ddagger$\end{example}

Before ending this section, a few comments can be brought. An extension of an ordinary differential system is not necessarily unique, as mentioned in the previous examples \ref{twri} and \ref{d-h}. For $N=1$ superspace (Grassmann-valued) supersymmetric extensions, one notices that so-called ``non-trivial'' extensions can be built of odd scalar superfields and suitable insertions of $D$  - derivatives of superfields in the case of polynomial expressions. Many evolution systems have been extended supersymmetrically while considering such aspects.

The operator $\epsilon Q$ acts as a derivation on scalar superfields.  In the case of systems extended using even scalar superfields, ``non-trivial'' systems might be exhibited using products involving $D$ - derivatives applied on superfields. To generate an extension using $N=1$ even superfields $X_i, i =1, ... , n$, terms of the type : $X_i, X_i X_j, DX_i DX_j, ...$, with $i,j = 1, ... , n$ can be explored. The parity, even or odd, of each equation of the built extended differential system being respected. As for the use of odd superfields $Y_i, i = 1, ... ,n$, terms of the form $Y_i, Y_iDY_j, Y_iDY_jDY_k$, with $i,j,k = 1, ... ,n$ can be attempted, with parity respected.

Integrability of a Grassmann-valued system is another aspect that could be considered, and it could constrain the availability of suitable extensions. First integrals of the non-extended system might be lost in carrying a supersymmetric extension, especially when odd superfields are at play.

\medskip\noindent
\section{\textit{N} - Extensions of Grassmann-valued systems}

Here, the even (or odd) superfield $X = X(t,\theta^1,\theta^2, ..., \theta^N)$, where $t \in$$^0\Lambda$ and $\theta^i \in$$^1\Lambda$, for each $i = 1, ..., N$, is defined on $\R^{(1,N)}$ superspace.

\begin{example}[$N=2$]
\begin{equation}
\dot X = F(X)
\end{equation}
where $N=2$ and the even superfield $X(t,\theta^1,\theta^2) = x(t) + \theta^1 \xi_1(t) + \theta^2 \xi_2(t) + \theta^1\theta^2 \chi(t)$,
where $x(t), \chi(t) \in$$^0\Lambda$, $\xi_1, \xi_2 \in$$^1\Lambda$, and $F(X)$ is a polynomial function of finite degree $n$ of $X$.
In general, multinomial series can be used for finite $N$ and $n$ of respectively $X$ and $F$ to retrieve explicit expressions for the polynomial $F(X)$, in terms of component fields. Thus for $N=2$, the highest degree term $X^n$ ($n \geq 2$) of $F(X)$ would have the form :
\begin{align}
X^n &= x^n + \theta^1 n x^{n-1} \xi_1 + \theta^2 nx^{n-1}\xi_2 + \theta^1\theta^2 n x^{n-1} \chi - \theta^1\theta^2 n(n-1) x^{n-2} \xi_1\xi_2 \\ \nonumber
X^n &= F(x) + \frac{\p F(x)}{\p x} (\theta^1 \xi_1 + \theta^2 \xi_2 + \theta^1\theta^2 \chi) - \theta^1\theta^2 \frac{\p^2F(x)}{\p x^2} \xi_1\xi_2
\end{align}

\hfill $\ddagger$\end{example}

For an odd superfield $X$, $X^n = 0$, for $n > 1$. Expressions can be derived for $N$ - extended even scalar superfields using the multinomial series (see for example \cite{Wei}). For instance, a $N$ - even superfield can be expanded as the sum of even $2^N$ terms. Recall the multinomial series :
\begin{equation}
(z_1 + z_2 + ... + z_m)^n = \sum_{k_1 + k_2 + ... + k_m = n}\binom{n}{k_1,k_2, ... , k_m} \prod_{i=1}^m (z_i)^{k_i},
\end{equation}
where the multinomial coefficient reads : $\binom{n}{k_1,k_2, ... , k_m} =\frac{n!}{k_1! k_2! ... k_m!}$.
Thus one can write : 
\begin{equation}
X = x + \sum_{j=1}^N\theta^j \xi_j + \sum_{i<j =1}^N \theta^i\theta^j \chi_{ij} + ... + \theta^1\theta^2 ... \theta^N \chi,
\end{equation}
and use $F(X) = X^n$, where $z_1 = x, z_2 = \theta^1 \xi_1, ... , z_{N+1} = \theta^N \xi_N, ... , z_m = \theta^1\theta^2 ... \theta^N \chi$, and $m = 2^N$. Let $n \geq N$, one is aware that among the sets $(k_1,k_2, ... ,k_m)$, where : $k_1 \leq n, k_2 \leq 1, ... , k_m \leq 1$, with $k_1 + k_2 + ... + k_m = n$, non-vanishing terms in the series could be obtained. When  $k_1 = n, k_2 = 0, ... , k_m = 0$, the multinomial coefficient $1$ is associated to the term $x^n = F(x)$. In the case when  $k_1 = n - 1, k_2 = 1, k_3 = 0, ... , k_m = 0$, a factor $n x^{n-1} = \frac{\p F(x)}{\p x}$ occurs. For  $k_1 = n - 2, k_2 = 1, k_3 = 1, k_4 = 0, ... , k_m = 0$, a factor $n(n-1) x^{n-2} = \frac{\p^2 F(x)}{\p x^2}$ appears. The contribution from the product of all the $N$ $\theta^i$s, with $k_1 = n - N, k_2 = 1, ... , k_{1+N} = 1, k_{2+N} = 0, ... , k_m = 0$ leads to a factor $\frac{n!}{(n-N)!} x^{n-N} = \frac{\p^N F(x)}{\p x^N}$. Further terms to the expansion can be derived in a similar manner.

For $\R^{(1,N)}$, with as above, coordinates $(t,\theta^1,\theta^2, ... ,\theta^N)$, one can define a supersymmetry operator :
\begin{equation}
Q  = \left (\frac{\p}{\p \theta^1} + \frac{\p}{\p \theta^2} + ... + \frac{\p}{\p \theta^N} + (\theta^1 + \theta^2 + ... + \theta^N)\frac{\p}{\p t} \right )
\end{equation}
and an operator :
\begin{equation}
D  = \left (\frac{\p}{\p \theta^1} + \frac{\p}{\p \theta^2} + ... + \frac{\p}{\p \theta^N} - (\theta^1 + \theta^2 + ... + \theta^N)\frac{\p}{\p t} \right )
\end{equation}
which satisfy : $\{Q,Q\} = 2N\frac{\p}{\p t} = - \{D,D\}$, as well as : $\{Q,D\} = 0$. The following transformations :\newline $Q_i = \left ( \frac{\p}{\p \theta^i} + \theta^i \frac{\p}{\p t} \right )$, and operators :  $D_i = \left ( \frac{\p}{\p \theta^i} - \theta^i \frac{\p}{\p t} \right )$, for $i = 1, ... ,N$, obey the anti-commutation relations : \newline $\{Q_i,Q_j\} = 2 \delta_{ij} \frac{\p}{\p t}$,  $\{D_i,D_j\} = - 2 \delta_{ij} \frac{\p}{\p t}$, and  $\{Q_i,D_j\} = 0$, for all $i,j = 1, ... , N$.

One can then ponder about a relatively general expectation. Let us consider a general equation for a ($N$ - extended) scalar superfield $X$ :
\begin{equation}
\mathcal{F}(X,\dot X, ... , X^{(n)}, D_i X, D_i D_j X, ...) = 0, \quad i,j = 1, ... , N
\end{equation}
Under conditions on the functional $\mathcal{F}$ of being well behaved with respect to derivatives  (for example, with respect to the chain rule, see for instance Thm 4.4.2 of \cite{Rog}), one might find $\mathcal{F} =  0$ to be invariant under the supersymmetry transformations : $\delta X = \epsilon Q X$.

\begin{example}[Polynomial terms and N=2 operators]
Let us indicate that the following simple system :
\begin{equation}
\dot X  = X^2 + (D_2X) (D_1 X),
\end{equation}
with $N=2$ even superfield $X =x + \theta^1 \xi_1 + \theta^2 \xi_2 + \theta^1\theta^2 \chi$, can be seen as a supersymmetric extension of $\dot \chi  = - (\chi)^2$. It gives in component form the set of equations :
\begin{align}
\dot x &= x^2 - \xi_1\xi_2 \\ \nonumber
\dot \xi_1 &= 2x \xi_1 + \xi_2 \dot x - \chi \xi_1\\ \nonumber
\dot \xi_2 &= 2x\xi_2 - \xi_2 \chi - \dot x \xi_1\\ 
\dot \chi &= 2x\chi - 2\xi_1\xi_2 - \xi_1\dot\xi_1 - \xi_2 \dot\xi_2 - (\chi)^2 - (\dot x)^2 \nonumber
\end{align}
One notes that the equation for the components $\chi$ and $x$ have quadratic nonlinearities even at the ``body'' or zero level of a Grassmann variable expansion. This non-trivial supersymmetric extension of $\dot \chi  = - (\chi)^2$ has the following supersymmetries : $\delta X  = \epsilon Q X$ (as well as $\delta X_1 = \epsilon^1 Q_1 X$ and  $\delta X_2 = \epsilon^2 Q_2 X$ supersymmetry transformations, where $\epsilon^1, \epsilon^2 \in$$^1\Lambda$).
\hfill $\ddagger$\end{example}

\begin{example}[N=2 Extensions of the H\'{e}non - Heiles system]
A first version is attained by simply embedding the variables $x^1$ and $x^3$ of a second-order version of the system, respectively in two even scalar superfields : $X_1(t,\theta^1,\theta^2) = x^1 + \theta^1 \xi_1^1 + \theta^2 \xi_2^1 + \theta^1\theta^2 \chi^1$ and $X_3(t,\theta^1,\theta^2) = x^3 + \theta^1 \xi_1^3 + \theta^2 \xi_2^3 + \theta^1\theta^2 \chi^3$. One writes an extension of the H\'enon - Heiles \cite{Hen} systems as :
\begin{equation}\label{henon-heiles}
\ddot X_1 = -X_1 - 2\lambda X_1 X_3, \quad \ddot X_3 = - X_3 - \lambda ((X_1)^2 - (X_3)^2),
\end{equation}
where $\lambda$ is a real constant, with invariance under supersymmetry transformations : $\delta X_1 = \epsilon Q X_1$, and $\delta X_3 = \epsilon Q X_3$. Note that a first order differential system extension can be obtained with $\dot X_1 = X_2$ and $\dot X_3 = X_4$, using similar expansions for $N=2$ superfields of the $x^2$ and $x^4$ variables. The above extension (\ref{henon-heiles}) could be labeled as trivial, but one could instead consider the following extension of the non-extended system given in terms of the $\chi^1$ and $\chi^3$ variables :
\begin{align}
\ddot X_1 &= - X_1 - 2 \lambda (D_1 X_1)(D_2 X_3)\\
\ddot X_3 &= - X_3 - \lambda ((D_1 X_1)(D_2 X_1) - (D_1 X_3)(D_2 X_3)),\nonumber
\end{align}
which leads to the component equations :
\begin{align}
\nonumber
\ddot x^1 &= - x^1 - 2 \lambda \xi_1^1 \xi_2^3\\ \nonumber
\ddot \xi_1^1 &= - \xi_1^1 - 2 \lambda(\xi_1^1 \chi^3 - \dot x^1 \xi_2^3)\\
\ddot \xi_2^1 &= - \xi_2^1 - 2 \lambda(\xi_1^1 \dot x^3 +  \chi^1 \xi_2^3)\\ \nonumber
\ddot \chi^1 &= - \chi^1 - 2 \lambda(\xi_1^1 \dot\xi_1^3 + \chi^1\chi^3 + \dot x^1\dot x^3 - \dot \xi_2^1 \xi_2^3)
\end{align}
and
\begin{align}
\nonumber
\ddot x^3 &= -x^3 - \lambda (\xi_1^1\xi_2^1 -  \xi_1^3 \xi_2^3) \\ \nonumber
\ddot \xi_1^3 &= - \xi_1^3 - \lambda [(\xi_1^1 \chi^1 - \dot x^1 \xi_2^1) - (\xi_1^3 \chi^3 - \dot x^3 \xi_2^3)]\\
\ddot \xi_2^3 &= - \xi_2^3 - \lambda [(\xi_1^1 \dot x^1 + \chi^1 \xi_2^1) - (\xi_1^3 \dot x^3 + \chi^3 \xi_2^3)]\\ \nonumber
\ddot \chi^3 &= - \chi^3 - \lambda [(\xi_1^1 \dot \xi_1^1 +(\chi^1)^2 + (\dot x^1)^2 - \dot \xi_2^1 \xi_2^1) \\ \nonumber
&\hskip 45pt - (\xi_1^3 \dot \xi_1^3 +(\chi^3)^2 + (\dot x^3)^2 - \dot \xi_2^3 \xi_2^3)] \nonumber
\end{align}
invariant under the supersymmetry transformations $\delta X_1 = \epsilon Q X_1$, and \newline 
$\delta X_3 = \epsilon Q X_3$. The H\'enon - Heiles system for $\chi^1, \chi^3$ is recovered by setting the other variables to zero. One might also be curious to explore for the existence of chaotic behaviour as the  H\'enon - Heiles system is itself embedded in each of the extensions, when components are explicitly expanded in terms of a Grassmann algebra basis.
\hfill $\ddagger$\end{example}

\begin{example}[N=3 extension of the Euler-Arnold equations]

The Euler - Arnold equations are related to geodesic motion on the Lie group $SO(n)$ given a left-invariant metric $\lambda$. Explicitly, in terms of elements $x^{ij}, i,j = 1, ... ,n$ of the Lie algebra $so(n)$ \cite{Hai}, one has :
\begin{equation}\label{euler-arnold}
\dot x^{ij} = \sum_{k=1}^n A_{kj}^{ki} x^{ik} x^{kj}
\end{equation}
where $A^{kl}_{ij} = \lambda_{ij} - \lambda_{kl}$. The Euler's equations for the torque free motion of a rigid body in three dimensions ($n=3$) can be retrieved with (symmetric) $\lambda_{ij} = \frac{1}{b_i+b_j}$, where the $b_i$ are constants, with $ i,j = 1, ..., n$.

Let the $N=3$ even scalar superfields be defined as : \newline $X^{ij}(t) = x^{ij}(t) + \sum_{l=1}^3 \theta^l \xi_l^{ij}(t) +\sum_{l<m = 1}^3 \theta^l\theta^m \chi_{lm}^{ij}(t) + \theta^1\theta^2\theta^3 \chi^{ij}(t), i,j = 1,2, ... ,n$, where all functions are antisymmetric with respect to the exchange of the indices $i$ and $j$, such that they belong to the Lie algebra $so(n)$. One can, analogously to the previous example, write an extension to the system (\ref{euler-arnold}) as :
\begin{equation}
\dot X^{ij} = \sum_{k = 1}^n A_{kj}^{ki} (D_1 X^{ik})(D_2 X^{kj}),
\end{equation}
which is invariant under the supersymmetry transformations : $\delta X^{ij} = \epsilon Q X^{ij}$, $i,j = 1, ... ,n$. The equations for the variables $\chi_{12}^{ij}, i = 1, ... ,n$, embed the Euler - Arnold system. Higher $N$ supersymmetric extensions are expected to be devised similarly.\hfill $\ddagger$
\end{example}

This section closes by pointing out that trivial $N$ - extensions can be obtained by simple substitutions of even $N$ -  superfields, but that nontrivial supersymmetric versions can be devised by focusing on field components found at higher degree in Taylor's expansions of superfields. Their occurrences as main fields can normally be brought through $D$ - derivatives of superfields.  

\medskip\noindent

\section{Comments on certain methods of solution}

First, one can mention the method of symmetry reduction, extended from its use in non-extended systems, generalized and applied to many supersymmetric systems (see for example \cite{AH,GH,BGH} and references therein), as well as the Hirota approach (see for instance \cite{DH1,DH2}) extended and applied to a set of supersymmetric systems. These methods have been used for supersymmetric generalizations of partial differential equations (for instance, evolution equations). Other methods have also been discussed (see \cite{Ci,Kh}). It is wished here to mention and comment on certain methods that could be involved to bring solutions to ordinary differential systems involving Grassmann variables, in particular, to systems that would also show invariance under supersymmetry transformations. Main elements and or steps are recalled below.

\subsection{Layer-by-layer description and iteration method}

In this approach, one solves a Grassmann-valued differential system at a level (or layer) of the Grassmann basis and then proceeds to seek solutions at the next level (or layer). This method has been mentioned and used for instance in the article of R. Heumann and N.S. Manton \cite{HM}.

One begins by expressing the Grassmann-valued variables in terms of a basis of a Grassmann algebra generated by $\{e^i, i = 1, ..., L (\text{finite or}\; \infty)\}$. This could be associated with ``bosonization'' for some authors (see for example \cite{GL,R}). For a scalar variable $V(t)$ belonging either to $^0\Lambda$ or $^1\Lambda$  , one has :
\begin{equation}
V(t) = V_0(t)\cdot 1 +\sum_{k=1, i_1<i_2 ... < i_k}^L V_{i_1,i_2, ... ,i_k}(t) e^{i_1} e^{i_2} ... e^{i_k}
\end{equation}

A level is labeled by the index $k = 0,1,2, ... ,L$. Recall that an even variable will show only the term $k=0$ and terms with an even number of generators $e^i$, an odd variable will be associated only with terms having an odd number of generators $e^i$. The set of variables at the $k=0$ level (or body variables) being denoted by $\vec V_0$, and at a level $k$, by $\vec V_k(t)$. A Grassmann-valued first order differential system could then be expressed as :
$\dot{\vec{V_0}} =\vec{f_0}(\vec{V_0})$, ... ,$\dot{\vec{V_k}} = \vec{f_k}(\vec{V_0}, ... , \vec{V_k})$ with $k=1, ... , L$.

As far as a system of first order differential equations is concerned, the known iteration method (see for instance \cite{CL}) could be attempted on the systems for $\vec V_k, k = 0,1,2, ... ,L$, in order to obtain solutions, given initial conditions, where the independent variable $t \in \R$, or a suitable interval. Thus, for a system : $\dot X = F(t,X)$, with scalar superfield $X(t) =  x(t) + \theta \xi(t)$, and initial condition $X(t_0)$, the $a^{th}$ iteration (successive approximation), denoted $X_a(t)$, would succinctly read :
\begin{equation}
X_a(t) = X(t_0) + \int_{t_0}^t F(\tilde t, X_{a-1}(\tilde t)) \;d\tilde t,
\end{equation}
where $a = 1,2,3, ...$. 
For existence and uniqueness results, one could consult \cite{CL}.

\begin{example}
For example, considering the simple system : $\dot X = F(t,X) = X^2$ with initial condition $X(t_0 = 0) = 1 + \theta \alpha$, where the constant $\alpha \in$$^1\Lambda$. One finds the second successive approximation :
\begin{equation}
X_2(t) = (1 + \theta \alpha) + \left (t + t^2 + \frac{t^3}{3} \right ) + \theta \alpha \left (2t + 3t^2 + \frac{4t^3}{3} \right ),
\end{equation}
which agrees with a solution to order $t^2$ obtained by solving more ``directly'' the system.

\hfill $\ddagger$
\end{example}

\subsection{Supersymmetry transformations}

Supersymmetry transformations might be useful to generate solutions of a supersymmetric system, with a nontrivial solution already known. For example, by inserting an even solution (at zero or body level) into supersymmetry transformations. Multiple successive supersymmetry transformations can be applied. This follows also for the case of $N$ extended supersymmetries.

\begin{example}
A simple example involves again the system : $\dot X = X^2$, where the scalar superfield $X = x + \theta \xi$, and its transformed scalar superfield has the form : $\tilde X = \tilde x + \theta \tilde \xi = X + \epsilon QX$. Using a supersymmetry transformation, the solution $x(t) =- \frac{1}{t}, \xi(t) = 0$, allows to write a different solution $\tilde X = -\frac{1}{t} + \theta (-)\frac{\epsilon}{t^2}$ to this system.

\hfill $\ddagger$
\end{example}

\subsection{Polynomial systems and non-associative products}

Some systems could be associated with commutative products with non-associative product aspects. A first - order system of the form :
\begin{equation}
\dot X_i = \mathcal{P}_i(X_j),
\end{equation}
where $i,j = 1, ... ,n$, $X_i$ are scalar superfields, and $\mathcal{P}_i$ are polynomials in $X_j$, can be transformed (enlarged) into a homogeneous system with the introduction of a new variable $u$; this by adding, if needed, a suitable power of this new variable to each term of the polynomials. It is mentioned in for example \cite{Wal,KS} that a (non-Grassmannian) system with homogeneous polynomials of degree $n$ can be rewritten as a system of homogeneous polynomials of degree 2 by adding a set of new variables. Putting $u=1$ in a solution of the homogenized system will provide a solution to the original system. Two examples are presented below where non-associative (not-associative)  product descriptions are exhibited. Series solutions can be probed.

\begin{example}\label{simple-nonlin-na}
The simple nonlinear system $\dot X = X + X^2$, where $X =  x + \theta \xi$ is an even scalar superfield is discussed below. It already involves a polynomial of degree 2, and a homogenization  can be obtained as : 
\begin{equation}
\bmatrix \dot x \\ \dot \xi \\ \dot u \endbmatrix = \bmatrix u x + x^2 \\ u\xi + 2 x \xi \\ 0\endbmatrix
\end{equation}
where $u \in$$^0\Lambda$. Defining the homogeneous quadratic mapping  $Q([X,u]) = [ux +x^2, u\xi + 2 x \xi, 0]^T$ (where $[...]^T$ stands for the transpose), one derives a bilinear product, denoted $B$ :
\begin{align}\label{na-product-sim}
B([X,u];[Y,v]) &= \frac{1}{2}\left [ Q([X,u] + [Y,v]) - Q([X,u]) - Q([Y,v]) \right ] \\
B([X,u];[Y,v]) &= \frac{1}{2} \bmatrix uy +vx + 2xy \\ u \chi + v \xi + 2x \chi + 2 y \xi \\ 0 \endbmatrix,
\end{align}
where $X =[x,\xi]$ and $Y = [y,\chi]$. It can be verified that the product $B$ is commutative and not associative, that is, respectively : $B([X,u];[Y,v]) = B([Y,v];[X,u])$ for all $[X,u],[Y,v]$, and \newline $B(B([X,u];[Y,v]);[Z,w]) \neq B([X,u];B([Y,v];[Z,w]))$, for at least some set(s) of $[X,u], [Y,v], [Z,w]$. 
\hfill $\ddagger$
\end{example}

\begin{example}\label{kdv-na}
 The following first-order differential system associated to a supersymmetric extension of the KdV equation (see example \ref{first-order-kdv}) is considered :
 \begin{equation}
 \dot X_1 = X_2, \quad \dot X_2 = X_3, \quad \dot X_3 = 6 X_1 X_2,
 \end{equation}
 where each $X_i, i=1,2,3$, stands for an even scalar superfield. This polynomial system of degree 2 can be homogenized by suitably inserting the variable $u \in$$^0\Lambda$ as follows :
 \begin{equation}
 \bmatrix  \dot X_1 \\ \dot X_2 \\ \dot X_3 \\ \dot u \endbmatrix = \bmatrix u X_2 \\ u X_3 \\ 6 X_1 X_2 \\ 0 \endbmatrix
 \end{equation}
 
 One can define a homogeneous quadratic form : \newline $Q(X)  =[u X_2,u X_3,6X_1X_2,0]^T$, where $X = [X_1,X_2,X_3,u]^T$. A commutative not-associative product, denoted $B$, is obtained with :
 \begin{align}\label{na-product-kdv}
 B(X;Y) &= \frac{1}{2} \left [Q(X+Y) - Q(X) - Q(Y) \right ] \\
 B(X;Y) &= \frac{1}{2} \bmatrix u Y_2 + v X_2 \\ u Y_3 + v X_3 \\ 6(X_1 Y_2 + Y_1 X_2) \\ 0 \endbmatrix,
 \end{align}
 where $X = [X_1,X_2,X_3,u]^T$, and $Y = [Y_1,Y_2,Y_3,v]^T$.
 It can be shown that the product $B$ is commutative : $B(X;Y) = B(Y;X)$ for all $X,Y$, and not associative : $B(B(X;Y);Z) \neq B(X;B(Y;Z))$ for at least some set(s) of $X,Y,Z$.
 \hfill $\ddagger$
\end{example}
 
 Details about expressing series solutions on certain intervals to the homogenized systems in non-Grassmannian setting can be found in \cite{Wal,KS}.
 
Reference \cite{KS} brings for example a series (Taylor) expansion that be generalized to $[X,u]$ of example \ref{simple-nonlin-na} or $X$ of example \ref{kdv-na}. (Maclaurin series can as well be set up.)  Let us first consider a Grassmann setting.
Starting with a Grassmann-valued function of $t$ : $\fraX(t) \in$$^0\Lambda$, and using a basis for the Grassmann algebra : $\{e^i, i = 1,2,...\}$, one can write a series expression :
\begin{equation}
\fraX(t) = x(t) +\sum_{i<j} \xi_{ij}(t) e^ie^j + ...
\end{equation}
with a Taylor expansion of each coefficient (ordinarily well behaved function of $t$) such as :
\begin{align}
x(t) &= x(0) +  \dot{x}(0) t +  \ddot{x}(0) \frac{t^2}{2!} + ...,\\
\xi_{ij}(t) &= \xi_{ij}(0) + \dot{\xi_{ij}}(0) t + \ddot{\xi_{ij}}(0)\frac{t^2}{2!} + ... \nonumber
\end{align}
It then follows that :
\begin{equation}
\fraX (t) = \fraX (0) + \dot{\fraX} (0) t + \ddot{\fraX}(0) \frac{t^2}{2!} + ...
\end{equation}
It is noted that the Leibniz rule applies to both non-associative products $B$ (\ref{na-product-sim}) and (\ref{na-product-kdv}). In the latter case (equation (\ref{na-product-kdv})) :
\begin{equation}
\dot{B}(X,Y) = B(\dot{X},Y) + B(X,\dot{Y})
\end{equation}

In the example \ref{kdv-na}, using the o.d.e.s, one finds that (see \cite{KS}) :
\begin{equation}
X(t) = X + X^2 t + X^3 t^2 + (X^2 X^2 + 2 X^4)\frac{t^3}{3} + ...
\end{equation}
where : as initial condition, $X = X(0)$, with the short notation $B(X,Y) = XY$, and the $n$th power notation $X^{n} = X X^{n-1}$. Convergence of such series would have to be analyzed. 

\medskip\noindent
\section{Conclusion}

In this article, notes and results were offered regarding generalizations of some ordinary differential equations and systems of ordinary differential equations, mainly of first - order, to Grassmann variables. These extended systems often exhibited invariance under supersymmetry transformations. In order to bring supersymmetric extensions, use has been made of superfields, either even or odd, defined on different superspaces. Various examples of supersymmetric extensions have been offered, including the cases of the three-wave resonant interactions and Euler - Arnold equations.

Some Darboux theory notions were discussed and inspired certain aspects (extensions) with respect to a formulation including Grassmann variables. Comments on some methods of solution of Grassmann-valued systems were also added, such as on an adapted non-associative product description for Grassmann-valued systems of equations.

Research directions that could be explored in the future include a larger probe of Darboux theory concepts, a study of integrability of certain systems, as well as a search of solutions or qualitative aspects for certain supersymmetric versions of systems of ordinary differential equations, in particular, for systems extended from systems showing non-integrable characteristics.
    

\end{document}